\documentclass{amsproc}

\theoremstyle{definition}

\theoremstyle{remark}

\numberwithin{equation}{section}

\begin{document}

\title{Integral pentagon relations for 3d superconformal indices}

\author{Ilmar Gahramanov}
\address{Institut f\"{u}r Physik und IRIS Adlershof, Humboldt-Universit\"{a}t zu Berlin,
Zum Grossen Windkanal 6, D12489 Berlin, Germany and Institute of Radiation Problems ANAS, B.Vahabzade 9, AZ1143 Baku, Azerbaijan}
\email{ilmar@physik.hu-berlin.de}
\thanks{The first author is grateful to Tudor Dimofte and Grigory Vartanov for useful discussions. The author also is grateful to Bruno Le Floch for helpful comments and for pointing out some misprints. The author would like to thank the organizers of the String--Math 2014 conference held at the University of Alberta, Edmonton, Canada, June 9-13, 2014 for the chance to present these results at the conference. The author also would like to thank Nesin Mathematics Village (Izmir, Turkey), the Chalmers University of Technology (Gothenburg, Sweden), Perimeter Institute for Theoretical Physics (Waterloo, Canada) and  the Institut des Hautes Etudes Scientifiques (Bures-sur-Yvette, France) for hospitality during his stay, where some parts of this work was done. The author was supported by the European Science Foundation through the research grant ITGP Short Visit 6454 during a visit to the Chalmers University of Technology and by the support of the Marie Curie International Research Staff Exchange Network UNIFY of the European Union’s Seventh Framework Programme [FP7-People-2010-IRSES] under grant agreement No 269217 during a visit to the Perimeter Institute.}

\author{Hjalmar Rosengren}
\address{Department of Mathematical Sciences, Chalmers University of Technology and University of Gothenburg, SE-412~96 G\"{o}teborg, Sweden}
\email{hjalmar@chalmers.se}

\date{January 1, 1994 and, in revised form, June 22, 1994.}



\begin{abstract}
The superconformal index of a three-dimensional supersymmetric field theory can be expressed in terms of basic hypergeometric integrals. By comparing the indices of dual theories, one can find new integral identities for basic hypergeometric integrals. Some of these integral identities have the form of the pentagon identity which can be interpreted as the 2--3 Pachner move for triangulated 3-manifolds.
\end{abstract}

\maketitle

\section{Introduction}
The superconformal index is one of the efficient tools in the study of non-perturbative aspects of supersymmetric field theory providing the most rigorous mathematical check of supersymmetric dualities. Recent progress in superconformal index computations have significant implications for mathematics. 

A rather striking example is the observation made by Dolan and Osborn \cite{Dolan:2008qi} that the superconformal index of four-dimensional theories is expressible in terms of elliptic hypergeometric integrals. Such integrals are a new class of special functions  \cite{Ros1,Rains,Spir} and they are of interest both in mathematics and in physics. The identification of superconformal indices of supersymmetric dual theories is given by the Weyl group symmetry transformations for certain elliptic hypergeometric functions on different root systems. The computations of the superconformal indices of supersymmetric dual theories in four dimensions have led to new non-trivial integral identities for elliptic hypergeometric functions \cite{Spiridonov:2009za, Spiridonov:2011hf, Sudano:2011aa, Kutasov:2014wwa}. 

Superconformal indices of three-dimensio\-nal theories have attracted much attention recently \cite{Hwang:2011qt,Kapustin:2011vz,Gahramanov:2013rda,Krattenthaler:2011da,Gahramanov:2013xsa}. Their exact computation yields new powerful verifications of various supersymmetric dualities as mirror symmetry, Seiberg-like duality etc. Since a three-dimensional superconformal index can be expressed in terms of basic hypergeometric integrals \cite{Gasper, Gasper2}, by studying supersymmetric dualities one can get new identities for this type of special functions. In this work we consider a special type of such identities, namely five term relations or the so-called pentagon identities. The pentagon relations are interesting from different aspects, see, for instance, \cite{Faddeev:1993rs, Gahramanov:2013rda, Kashaev:2014rea, Kashaev:2012cz}. Here we present some examples of integral pentagon relations related to the three-dimensional superconformal index.

The rest of the paper is organized in the following way. In Section 2 we make a brief review of the superconformal index in three dimensions. We present some examples of the integral pentagon identities in Sections 3 and 4 and briefly discuss some open problems in Section 5.

\section{The superconformal index}
In this section we give a short introduction to a three-dimensional superconformal index and refer the reader to \cite{Imamura:2011su, Krattenthaler:2011da}  and references therein for more details.

Before proceeding, it would be useful to recall the well-known Witten index. Consider a supersymmetric quantum mechanics
\begin{align} 
	  & \{Q,Q^{\dag}\} \  = \ 2{ H} \;, \\
	  & \{Q,(-1)^F\} \ = \ 0 \;,
\end{align}
where $Q$, $H$ and $(-1)^F$ are the supersymmetric charge, the Hamiltonian and the fermion number operator\footnote{A fermion number $F$ takes the value zero on bosons and one on fermions.}, respectively. In order to check whether the supersymmetry is broken or not, Witten introduced the topological invariant of a theory \cite{Witten:1982df}
\begin{equation} \label{Wittenindex}
\text{I}_W=\text{Tr}(-1)^F e^{-\beta H} \;,
\end{equation}
which tells us that supersymmetry is not spontaneously broken if $\text{I}_W \neq 0$. The sum in the definition (\ref{Wittenindex}) runs over all physical states of the theory. The Witten index is independent of the parameter $\beta$ and counts the difference between the number of bosonic and fermionic ground states. It is an analogue of the Atiyah--Singer index \cite{Atiyah}. 

In the case of a supersymmetric field theory one can generalize the Witten index\footnote{The original Witten index for supersymmetric gauge theories gives the dual Coxeter number for the gauge group.} by including to the index global symmetries of a theory commuting with $Q$ and $Q^{\dagger}$ \cite{Kinney:2005ej, Romelsberger:2005eg, Romelsberger:2007ec}. For a $d$-dimensional supersymmetric theory the superconformal index is the following partition function defined on $S^{d-1} \times S^1$,
\begin{equation}
\text{I}(\{t_i\})=\text{Tr} (-1)^F e^{-\beta \{ Q, Q^{\dagger}\}} \prod t_i^{F_i} \;,
\end{equation}
where the trace is taken over the Hilbert space on $S^{d-1}$, $F_i$ are generators for global symmetries that commute with $Q$ and $Q^{\dagger}$, and $t_i$ are additional regulators (fugacities) corresponding to the global symmetries. The superconformal index counts short BPS operators of the theory. We refer the reader to \cite{Romelsberger:2007ec, Spiridonov:2009za, Gahramanov:gka} for more details and references.

We will restrict ourselves to three-dimensional theories with ${\mathcal N}=2$ supersymmetry\footnote{There are many interesting results in this direction also for theories with ${\mathcal N}=4$ \cite{Kapustin:2011jm}, ${\mathcal N}=6$ \cite{Bhattacharya:2008bja,Kim:2009wb} and ${\mathcal N}=8$ \cite{Bashkirov:2011pt} supersymmetry.}. These theories have four real supercharges which we organize into complex spinor $Q_{\alpha}$ and its conjugate $\bar{Q}_{\alpha}$ with $\alpha=1,2$. We focus on theories in the far IR, where the symmetry is enhanced to the superconformal symmetry and therefore there are four additional supercharges $S_{\alpha}$ and $\bar{S}_{\alpha}$ with $\alpha=1,2$. We choose one of the supercharges, say, $Q=Q_1$. For the theory on $S^2 \times S^1$ we have $Q_1^{\dagger}=S_1$ and the following commutation relation (for the full superconformal algebra, see \cite{Dolan:2008vc})
\begin{equation} \label{Ham}
\{Q,Q^{\dagger}\}=\Delta-R-j_3 \;,
\end{equation}
where $\Delta$, $R$,  $j_3$ are the energy, the generator of the $R$-symmetry and the third component of the angular momentum on $S^2$, respectively. Note that, because of conformal symmetry, $R$-symmetry appears explicitly in the commutation relations\footnote{In the case of ${\mathcal N}=2$ supersymmetric theories without conformal symmetry, the $R$-symmetry is only an automorphism group and does not appear in a direct way as in (\ref{Ham}). For details see, for instance, \cite{Aharony:1997bx}.}. The superconformal index of this theory is defined as \cite{Bhattacharya:2008bja}
\begin{equation}
\text{Tr}\left[(-1)^F e^{-\beta \{Q,Q^{\dagger}\}} x^{\Delta+j_3}\prod_i t_i^{F_i}\right] \;,
\end{equation}
where  $F_i$ are generators for flavor symmetries of the theory. 

Using the localization technique \cite{Pestun:2007rz} it was shown that the superconformal index of  a three-dimensional theory has the form of matrix integral (see, for instance, \cite{Kim:2009wb, Imamura:2011su}) which is a basic hypergeometric integral\footnote{In the literature it is also called $q$-hypergeometric integral.} \cite{Gasper, Gasper2}. For instance, the chiral multiplet with components being a complex scalar field $\phi$, a Weyl fermion $\psi$ and an auxiliary complex scalar $F$,
\begin{equation}
\Phi=\phi+\sqrt{2} \theta \psi +\theta^2 F
\end{equation}
($\theta$ is a Grassman coordinate),
and with $R$-charge $r$ in the fundamental representation of the gauge group $U(N)$, contributes to the index as
\begin{equation} \label{chiral}
\prod_{a=1}^{N_c} \frac{(x^{2-r+|m_a|} z_a^{-1}; x^2)_{\infty}}{(x^{r+|m_a|} z_a; x^2)_{\infty}} \;,
\end{equation}
where the $q$-Pochhammer symbol is defined as 
\begin{equation}
(z;q)_\infty=\prod_{i=0}^\infty (1-z q^i) \;
\end{equation}
and the integer parameters $m_i$ stand for the magnetic charges corresponding to the gauge group $U(N)$ and run over integers; the fugacities $z_i$ correspond to the gauge group. 

One can also compute the superconformal index by using representations of the superconformal algebra \cite{Krattenthaler:2011da}, i.e. by the so-called Romelsberger prescription \cite{Romelsberger:2007ec}.

From now on we will use the notation
\begin{equation}
x^2=q,
\end{equation}
and express the superconformal index via the so-called tetrahedron index \cite{Dimofte:2011py}
\begin{equation} 
{\mathcal I}_q[m,z] \ = \ \prod_{i=0}^{\infty} \frac{1-q^{i-\frac12 m+1}z^{-1}}{1-q^{i-\frac12 m} z}, \;\;\; \text{with $|q|<1$ and $m \in Z$.}
\end{equation}
The contribution of a chiral multiplet in terms of tetrahedron index has the following form
\begin{equation}
 \prod_{i=0}^{\infty} \frac{1-q^{i+\frac12 |m|+1} z^{-1}} {1-q^{i+\frac12|m| }z} \ = \ (-q^{\frac12})^{-\frac12(m+|m|)} z^{\frac12(m+|m|)} {\mathcal I}_q[m,z].
\end{equation}
We defined the tetrahedron index for the free chiral with zero $R$-charge, but one can write the index for general $R$-charge by the shift $z \rightarrow z q^{r/2}$.

The three-dimensional index can be factorized into vertex and anti-vertex partition function \cite{Krattenthaler:2011da,Pasquetti:2011fj,Hwang:2012jh} and all results presented in the next section can be written in this fashion; however, this subject is beyond the scope of the present work.

It is worth to mention here that we will consider theories with $U(1)$ gauge group only in order to obtain pentagon identities which we will study in the next section.

\section{Integral pentagon identities}

Our main interest is the five-term relation for the superconformal index. Let us consider the  $d=3$ ${\mathcal N}=2$ supersymmetric quantum electrodynamics with $U(1)$ gauge group and one flavor. The superconformal index of this theory is
\begin{equation}  \label{mirrorsym1}
I_e \ = \  \sum_{m \in Z} \oint \frac{dz}{2\pi i z} z^{-m} \; {\mathcal I}_q[m;q^{1/6}z^{-1}] \; {\mathcal I}_q[-m;q^{1/6}z] \;,
\end{equation}
where the integration is over the unit circle with positive orientation. For simplicity we switched off\footnote{See, for instance, \cite{Krattenthaler:2011da, Kapustin:2011jm}. We consider the influence of the topological $U(1)_J$ symmetry to the index in the next chapter, where we define the so-called generalized superconformal index.} the topological symmetry $U(1)_J$. 

The dual theory is the free Wess--Zumino theory\footnote{In the literature this theory sometimes is called the XYZ model.} \cite{Intriligator:1996ex, deBoer:1997ka, Aharony:1997bx} with three chiral multiplets $q,\tilde{q}$, $S$ interacting through the superpotential\footnote{The permutation symmetry of the superpotential fixes the $R$-charges, but one can write the index for more general $R$-charge like in \cite{Imamura:2011su}.} $W=\tilde{q} S q$.  The index of this theory has more simple form, since we do not need to integrate over the gauge group,
\begin{equation} 
I_m \ = \ \left( {\mathcal I}_q[0;q^{1/3}]  \right)^3 \;.
\end{equation}
These two theories are dual under the mirror symmetry, i.e.\ under exchange of the Higgs and the Coulomb branches \footnote{In three-dimenisonal supersymmetric theories the Coulomb and the Higgs branch are both hyper-K\"ahler manifolds.}. The mirror duality leads to the following integral pentagon identity
\begin{equation} \label{firstpentagon}
 \sum_{m \in Z} \oint \frac{dz}{2\pi i z} z^{-m} \; {\mathcal I}_q[m;q^{1/6}z^{-1}] \; {\mathcal I}_q[-m;q^{1/6}z] \ = \ \left( {\mathcal I}_q[0;q^{1/3}]  \right)^3 \;. 
\end{equation}
This is the first example of a pentagon identity for the tetrahedron index. The identity (\ref{firstpentagon}) was obtained in \cite{Krattenthaler:2011da} by using the superconformal index technique. In \cite{Krattenthaler:2011da} the authors also presented the mathematical proof of the integral identity (\ref{firstpentagon} by using Ramanujan summation formula for basic hypergeometric series. 

The tetrahedron index can be written in the following form: 
\begin{equation}
{\mathcal I}_q[m,z] =\sum_{e \in Z} {\mathcal I}(m,e) z^e \;,
\end{equation}
where
\begin{equation}
{\mathcal I}(m,e) =\sum_{n=\frac{1}{2}(|e|-e)}^{\infty}\frac{(-1)^n q^{\frac12 n (n+1)-(n+\frac12 e)m}}{(q)_n (q)_{n+e}} \;.
\end{equation}
This index was introduced in \cite{Dimofte:2011py}. It is also interesting from a mathematical point of view \cite{Garoufalidis3d, Garoufalidis:2013axa}. The index ${\mathcal I}(m,e)$ obeys the following pentagon identity
\begin{multline} \label{DGGpentagon}
{\mathcal I}(m_1-e_2,e_1){\mathcal I}(m_2-e_1,e_2) \\
 = \ \sum_{e_3} q^{e_3} {\mathcal I}(m_1,e_1+e_3){\mathcal I}(m_2,e_2+e_3){\mathcal I}(m_1+m_2,e_3).
\end{multline}
The proof of the identity (\ref{DGGpentagon}) is given in the Appendix of \cite{Garoufalidis3d}. This pentagon relation is a counterpart of the integral pentagon identity (\ref{firstpentagon}). In order to distinguish between this type relation and the identity of the form (\ref{firstpentagon}) we use the terminology ``the \textit{integral} pentagon identity'' for the latter one. 

As another example, we consider the following three-dimensional duality. The electric theory is the $d=3$ ${\mathcal N}=2$ superconformal field theory with $U(1)$ gauge symmetry and six chiral multiplets, half of them transforming in the fundamental representation  and another half transforming in the anti-fundamental representation of the gauge group. Its mirror dual is a theory with nine chirals and without gauge degrees of freedom (the gauge symmetry is completely broken). The mirror symmetry leads to the following identity
\begin{equation} \label{prepentagon}
\sum_{m \in Z} \oint \frac{d z}{2 \pi i z} \; (-z)^{-3 m} \prod_{i=1}^{3} {\mathcal I}_q[-m,q^{\frac16}\xi_i z] \; {\mathcal I}_q[m,q^{\frac16}\eta_i z^{-1}] = \prod_{i,j=1}^3 {\mathcal I}_q[0,q^{\frac13}\xi_i \eta_j] \; ,
\end{equation}
where the fugacities $\xi_i$ and $\eta_i$ stand for the flavor symmetry $SU(3)\times SU(3)$ and there is the balancing condition $\prod_{i=1}^3 \xi_i = \prod_{i=1}^3  \eta_i =1$. Note that we again dropped the topological symmetry $U(1)_J$. The identity (\ref{prepentagon}) was introduced in \cite{Gahramanov:2013rda}, to where we refer the reader for the details and the mathematical proof of it. 

Following \cite{Gahramanov:2013rda} we introduce a new function
\begin{equation}
{\mathcal B}[m;a,b]=\frac{{\mathcal I}_q[m,a] \; {\mathcal I}_q[-m,b]}{{\mathcal I}_q[0,a b]} \; ,
\end{equation}
and rewrite the equality (\ref{prepentagon}) in terms of this function. The final result is a new integral pentagon identity in terms of ${\mathcal B}[m;a,b]$ functions
\begin{equation} \label{pentagon}
 \sum_{m\in Z} \oint \frac{d z}{2 \pi i z}  \; (-z)^{-3 m} \prod_{i=1}^3 {\mathcal B}[m; \xi_i z^{-1}, \eta_i z]= {\mathcal B}[0; \xi_1 \eta_2, \xi_3 \eta_1] \; {\mathcal B}[0; \xi_2 \eta_1,  \xi_3 \eta_2]
\end{equation}
where we have redefined the flavor fugacities  $\xi_i \rightarrow q^{-1/6} \xi_i$ and $\eta_i \rightarrow q^{-1/6} \eta_i$ and the new  balancing condition\footnote{We have a misprint in our previous paper \cite{Gahramanov:2013rda}.} is $\prod_{i=1}^3 \xi_i = \prod_{i=1}^3  \eta_i =q$.

\section{Generalized superconformal index}

One can also find a similar pentagon relation for the \textit{generalized superconformal index} \cite{Kapustin:2011jm}. Unlike four-dimensional gauge theories, in three dimensions there are no chiral anomalies, therefore there is no obstruction for considering a theory in a non-trivial background gauge field coupled to the global symmetries. Then one gets new discrete parameters for global symmetries in the expression of the superconformal index. The index with new integer parameters corresponding to the global symmetries is called the generalized superconformal index. 

One can apply the above techniques similarly to the generalized superconformal index and obtain more general integral pentagon identities.
 
The expression (\ref{prepentagon}) in terms of the generalized index has the following form
\begin{eqnarray}  \nonumber
\sum_{m \in Z}  \int_{\mathbb{T}} \frac{dz}{2\pi i z} \prod_{i=1}^3 (-1)^{m} z^{-3m}\frac{((\xi_i z)^{-1} q^{1+m/2};q)_{\infty}(z/\eta_i q^{1-m/2};q)_{\infty}}{(\xi_i z q^{m/2+{ M_i}};q)_{\infty}(\eta_i/z q^{-m/2+{ N_i}};q)_{\infty}} \\
 = \ \frac{1}{\prod_{j=1}^3 q^{\binom{M_j}2+\binom{N_j}2}\xi_j^{M_j}\eta_j^{N_j}} \prod_{i,j=1}^{3} \frac{(q/\xi_i \eta_j;q)_{\infty}}{(\xi_i \eta_j q^{{M_i+N_j}};q)_{\infty}}
\end{eqnarray}
where we switched on background fields coupled to the flavor symmetry and therefore the index has additional integer parameters $M_i$ and $N_i$ with the condition $\sum_i M_i=\sum_i N_i=0$. There is also the balancing condition $\prod_{i=1}^3 \xi_i = \prod_{i=1}^3  \eta_i =q$ for flavor fugacities. The new discrete parameters are analogous to the magnetic charge $m$ for the gauge symmetry. 

The analogue of the first pentagon identity (\ref{firstpentagon}) in terms of the generalized superconformal index is the following pentagon identity
\begin{eqnarray} \nonumber
\sum_{s \in Z} \int \frac{dz}{2 \pi i z}  z^{2n-s} \omega^m \alpha^{-m} q^{\frac14 m} {\mathcal I}_q[m+s;q^{\frac14}\alpha z^{-1}] {\mathcal I}_q[m-s; \alpha z q^{\frac14}]\\ \nonumber
=  \omega^{-m} \alpha^{n+2m} q^{\frac14 n} {\mathcal I}_q[m;q^{\frac14}\alpha^{-1} \omega^{-1}] {\mathcal I}_q[-m; q^{\frac14} \alpha^{-1} \omega ] {\mathcal I}_q[2m;q^{\frac12}\alpha^2],
\end{eqnarray}
where we switched on the background gauge field coupled to the topological $U(1)_J$ global symmetry. Here $\alpha$ and $m$ denote the parameters for the axial $U(1)_A$ symmetry, $\omega$ and $n$ denote the parameters for the topological $U(1)_J$ symmetry and the discrete parameter $s$ stands for magnetic charge. Note that this identity was proven only for the case $m=0$ \cite{Kapustin:2011jm}. 

\section{Concluding remarks}

There is a recently proposed relation called \textit{$3d-3d$ correspondence} \cite{Dimofte:2011py, Dimofte:2011ju} (see also \cite{Terashima:2011qi, Terashima:2011xe, Terashima:2012cx, Galakhov:2012hy, Dimofte:2013iv, Gang:2013sqa, Chung:2014qpa}) that connects $d=3$ ${\mathcal N}=2$ supersymmetric theories and triangulated 3-manifolds. Namely, the independence of the invariant of the corresponding 3-manifold on the choice of triangulation corresponds to the equality of superconformal indices of mirror dual theories \cite{Dimofte:2011py,Garoufalidis:2013axa}. In this context the interpretation of the integral pentagon identities discussed here is the 2--3 Pachner move \cite{Pachner1,Pachner2} for triangulated 3-manifolds, which relates different  decompositions of a polyhedron with five ideal vertices into ideal tetrahedra. Much work remains to be done in this direction. 

Note that one can write such pentagon identities also for partition functions on $S^3$ \cite{Kashaev:2012cz}, i.e.\ for hyperbolic hypergeometric integrals \cite{Bult}.  

As an aside comment, we would like to mention that the pentagon identity (\ref{pentagon}) may represent the star-triangle relation for some integrable model.

\bibliographystyle{amsalpha}

\begin{thebibliography}{A}

\bibitem[AHISS]{Aharony:1997bx}
O.~Aharony, A.~Hanany, K.~A. Intriligator, N.~Seiberg, and M.~Strassler,
  \textit{Aspects of N=2 supersymmetric gauge theories in three-dimensions},
  {{Nucl.Phys.}
  {\bfseries B499} (1997) 67--99},
{{\ttfamily arXiv:hep-th/9703110}}.

\bibitem[AS]{Atiyah}
M.~F. Atiyah and I.~M. Singer, \textit{The index of elliptic operators on compact
  manifolds}, { Bulletin of the American Mathematical Society} {\bfseries
  69} no.~3, (05, 1963) 422--433.




\bibitem[BK]{Bashkirov:2011pt}
  D.~Bashkirov and A.~Kapustin,
  \textit{Dualities between N = 8 superconformal field theories in three dimensions},
  JHEP {\bf 1105} (2011) 074
  {\ttfamily arXiv:1103.3548}.

\bibitem[BM]{Bhattacharya:2008bja}
J.~Bhattacharya and S.~Minwalla, \textit{Superconformal Indices for N = 6 Chern
  Simons Theories},
 {{ JHEP} {\bfseries
  0901} (2009) 014},
{{\ttfamily arXiv:0806.3251}}.

\bibitem[BHOY]{deBoer:1997ka}
J.~de~Boer, K.~Hori, Y.~Oz, and Z.~Yin, \textit{Branes and mirror symmetry in N=2 supersymmetric gauge theories in three-dimensions},
 {{Nucl.Phys.}
  {\bfseries B502} (1997) 107--124},
{{\ttfamily arXiv:hep-th/9702154}}.

\bibitem[Bu]{Bult}
F.~J. van~de Bult, \textit{Hyperbolic hypergeometric functions}, {PhD thesis}
  (2007) .


\bibitem[CDGS]{Chung:2014qpa}
H.-J. Chung, T.~Dimofte, S.~Gukov, and P.~Sulkowski, \textit{3d-3d Correspondence
  Revisited},
{{\ttfamily arXiv:1405.3663}}.



\bibitem[DGaGo]{Dimofte:2013iv}
T.~Dimofte, M.~Gabella, and A.~B. Goncharov, \textit{K-Decompositions and 3d Gauge
  Theories},
{{\ttfamily arXiv:1301.0192}}.

\bibitem[DGG1]{Dimofte:2011py}
T.~Dimofte, D.~Gaiotto, and S.~Gukov, \textit{3-Manifolds and 3d Indices},
{{\ttfamily arXiv:1112.5179}}.

\bibitem[DGG2]{Dimofte:2011ju}
T.~Dimofte, D.~Gaiotto, and S.~Gukov, \textit{Gauge Theories Labelled by
  Three-Manifolds}, {{Commun.Math.Phys.} {\bfseries 325} (2014) 367--419},
{{\ttfamily arXiv:1108.4389}}.

\bibitem[D]{Dolan:2008vc}
F.~Dolan, \textit{On Superconformal Characters and Partition Functions in Three
  Dimensions}, {{J.Math.Phys.}
  {\bfseries 51} (2010) 022301},
{{\ttfamily arXiv:0811.2740}}.

\bibitem[DO]{Dolan:2008qi} 
  F.~A.~Dolan and H.~Osborn,
  \textit{Applications of the Superconformal Index for Protected Operators and q-Hypergeometric Identities to N=1 Dual Theories},
  Nucl.\ Phys.\ B {\bf 818}, 137 (2009)
  {\ttfamily arXiv:0801.4947}.

\bibitem[FK]{Faddeev:1993rs}
L.~Faddeev and R.~Kashaev, \textit{Quantum Dilogarithm},
  {{ Mod.Phys.Lett.}
  {\bfseries A9} (1994) 427--434},
{{\ttfamily arXiv:hep-th/9310070}}.

\bibitem[GR]{Gahramanov:2013rda}
I.~Gahramanov and H.~Rosengren, \textit{A new pentagon identity for the tetrahedron
  index}, {{ JHEP}
  {\bfseries 1311} (2013) 128},
{{\ttfamily arXiv:1309.2195}}.

\bibitem[GV1]{Gahramanov:gka}
I.~B. Gahramanov and G.~S. Vartanov, \textit{Superconformal indices and partition
  functions for supersymmetric field theories},
 {{ XVIIth International
  Congress on Mathematical Physics} (2013) 695--703},
{{\ttfamily arXiv:1310.8507}}.

\bibitem[GV2]{Gahramanov:2013xsa} 
  I.~Gahramanov and G.~Vartanov,
  \textit{Extended global symmetries for 4D $N$ = 1 SQCD theories},
  J.\ Phys.\ A {\bf 46}, 285403 (2013)
  {\ttfamily arXiv:1303.1443}.

\bibitem[GMMS]{Galakhov:2012hy}
  D.~V.~Galakhov, A.~Mironov, A.~Morozov, A.~Smirnov,
  \textit{Three-dimensional extensions of the Alday-Gaiotto-Tachikawa relation},
  Theor.\ Math.\ Phys.\  {\bf 172} (2012) 939
   {\ttfamily arXiv:1104.2589}.

\bibitem[GKLP]{Gang:2013sqa}
D.~Gang, E.~Koh, S.~Lee, and J.~Park, \textit{Superconformal Index and 3d-3d
  Correspondence for Mapping Cylinder/Torus},
{{ JHEP} {\bfseries 1401}
  (2014) 063},
{{\ttfamily arXiv:1305.0937}}.


\bibitem[Gar]{Garoufalidis3d}
S.~Garoufalidis, \textit{The 3D index of an ideal triangulation and angle
  structures}, {{\ttfamily arXiv:1208.1663}}.


\bibitem[GHRS]{Garoufalidis:2013axa}
S.~Garoufalidis, C.~D. Hodgson, J.~H. Rubinstein, and H.~Segerman,
  \textit{1-efficient triangulations and the index of a cusped hyperbolic
  3-manifold},{{\ttfamily arXiv:1303.5278}}.

\bibitem[G]{Gasper}
G.~{Gasper}, ``{Lecture notes for an introductory minicourse on q-series},''
  {{\ttfamily
  arXiv:math/9509223}}.



\bibitem[GaRa]{Gasper2}
G.~{Gasper} and M.~Rahman, \textit{Basic Hypergeometric Series}, {Cambridge
  University Press, 2nd ed} (2004) .
  



\bibitem[HKP]{Hwang:2012jh} 
  C.~Hwang, H.~C.~Kim and J.~Park,
  \textit{Factorization of the 3d superconformal index}, JHEP {\bf 1408}, 018 (2014){\ttfamily arXiv:1211.6023}.

  
\bibitem[HKPP]{Hwang:2011qt} 
  C.~Hwang, H.~Kim, K.~J.~Park and J.~Park,
  \textit{Index computation for 3d Chern-Simons matter theory: test of Seiberg-like duality},
  JHEP {\bf 1109}, 037 (2011)
  {\ttfamily arXiv:1107.4942}.

\bibitem[IY]{Imamura:2011su}
Y.~Imamura and S.~Yokoyama, \textit{Index for three dimensional superconformal field
  theories with general R-charge assignments},
 {{ JHEP} {\bfseries 1104}
  (2011) 007},
{{\ttfamily arXiv:1101.0557}}.

\bibitem[IS]{Intriligator:1996ex}
K.~A. Intriligator and N.~Seiberg, \textit{Mirror symmetry in three-dimensional gauge theories}, {{Phys.Lett.} {\bfseries B387} (1996) 513--519},
{{\ttfamily arXiv:hep-th/9607207}}.

\bibitem[KKP]{Kapustin:2011vz} 
  A.~Kapustin, H.~Kim and J.~Park,
  \textit{Dualities for 3d Theories with Tensor Matter},
  JHEP {\bf 1112}, 087 (2011)
  {\ttfamily arXiv:1110.2547}.


\bibitem[KW]{Kapustin:2011jm}
A.~Kapustin and B.~Willett, \textit{Generalized Superconformal Index for Three
  Dimensional Field Theories},
{{\ttfamily arXiv:1106.2484}}.

\bibitem[K]{Kashaev:2014rea}
R.~Kashaev, \textit{On beta pentagon relations},
{{\ttfamily arXiv:1403.1298}}.

\bibitem[KLV]{Kashaev:2012cz} 
  R.~Kashaev, F.~Luo and G.~Vartanov,
  \textit{A TQFT of Turaev-Viro type on shaped triangulations},  {\ttfamily arXiv:1210.8393}.


\bibitem[Ki]{Kim:2009wb}
S.~Kim, \textit{The Complete superconformal index for N=6 Chern-Simons theory},
  {{ Nucl.Phys.} {\bfseries B821} (2009)
  241--284},
{{\ttfamily arXiv:0903.4172 [hep-th]}}.



\bibitem[KMMR]{Kinney:2005ej}
J.~Kinney, J.~M. Maldacena, S.~Minwalla, and S.~Raju, \textit{An Index for 4
  dimensional super conformal theories},
 {{Commun.Math.Phys.}
  {\bfseries 275} (2007) 209--254},
{{\ttfamily arXiv:hep-th/0510251}}.

\bibitem[KSV]{Krattenthaler:2011da}
C.~Krattenthaler, V.~Spiridonov, and G.~Vartanov, \textit{Superconformal indices of
  three-dimensional theories related by mirror symmetry},
 {{JHEP} {\bfseries 1106}
  (2011) 008},
{{\ttfamily arXiv:1103.4075}}.

\bibitem[KL]{Kutasov:2014wwa} 
  D.~Kutasov and J.~Lin,
  \textit{N=1 Duality and the Superconformal Index},
  {\ttfamily arXiv:1402.5411}.

\bibitem[Pa1]{Pachner1} U.~Pachner, \textit{Konstruktionsmethoden und das kombinatorische Hom\"{o}omorphieproblem f\"{u}r Triangulationen kompakter semilinearer Mannigfaltigkeiten}, Abh. Math. Sem.Univ. Hamburg 57 (1986) 69.

\bibitem[Pa2]{Pachner2}U.~Pachner, \textit{PL homeomorphic manifolds are equivalent by elementary shellings}, European J. Combin. 12 (1991) 129–145

\bibitem[Pas]{Pasquetti:2011fj} 
  S.~Pasquetti, \textit{Factorisation of N = 2 Theories on the Squashed 3-Sphere}, JHEP {\bf 1204}, 120 (2012)
  {\ttfamily arXiv:1111.6905}.

\bibitem[Pes]{Pestun:2007rz}
V.~Pestun, \textit{Localization of gauge theory on a four-sphere and supersymmetric
  Wilson loops}, {{Commun.Math.Phys.} {\bfseries 313} (2012) 71--129},
{{\ttfamily arXiv:0712.2824}}.


\bibitem[Ra]{Rains}
E. M. Rains, \textit{Transformations of elliptic hypergeometric integrals}, Ann.\ Math.\ {\bf 171} (2010), 169, {\ttfamily arXiv:math/0309252}.


\bibitem[R1]{Romelsberger:2005eg}
C.~Romelsberger, \textit{Counting chiral primaries in N = 1, d=4 superconformal
  field theories},
 {{Nucl.Phys.}
  {\bfseries B747} (2006) 329--353},
{{\ttfamily arXiv:hep-th/0510060}}.

\bibitem[R2]{Romelsberger:2007ec}
C.~Romelsberger, \textit{Calculating the Superconformal Index and Seiberg
  Duality},
{{\ttfamily arXiv:0707.3702}}.

\bibitem[Ro]{Ros1}
 H.\ Rosengren, \textit{Elliptic hypergeometric series on root systems}, Adv. Math. {\bf 181} (2004), 417, {\ttfamily arXiv:math/0207046}. 
  
   

\bibitem[Sp]{Spir}
 V.~P. Spiridonov, \textit{Essays on the theory of elliptic hypergeometric
   functions}, 
   Russ. Math. Surv. {\bfseries 63} no.~3, (2008) 405,
  {\ttfamily arXiv:0805.3135}.

\bibitem[SV1]{Spiridonov:2009za}
V.~Spiridonov and G.~Vartanov, \textit{Elliptic Hypergeometry of Supersymmetric
  Dualities}, {{
  Commun.Math.Phys.} {\bfseries 304} (2011) 797--874},
{{\ttfamily arXiv:0910.5944}}.

\bibitem[SV2]{Spiridonov:2011hf} 
  V.~P.~Spiridonov and G.~S.~Vartanov,
  \textit{Elliptic hypergeometry of supersymmetric dualities II. Orthogonal groups, knots, and vortices},
  Commun.\ Math.\ Phys.\  {\bf 325}, 421 (2014)
  {\ttfamily arXiv:1107.5788}.

 \bibitem[S]{Sudano:2011aa} 
   M.~Sudano,
   \textit{The Romelsberger Index, Berkooz Deconfinement, and Infinite Families of Seiberg Duals},
   JHEP {\bf 1205}, 051 (2012)
   {\ttfamily arXiv:1112.2996}.

\bibitem[TM1]{Terashima:2011qi}
Y.~Terashima and M.~Yamazaki, \textit{SL(2,R) Chern-Simons, Liouville, and Gauge
  Theory on Duality Walls},
  {{ JHEP} {\bfseries 1108}
  (2011) 135},
{{\ttfamily arXiv:1103.5748}}.

\bibitem[TM2]{Terashima:2011xe}
Y.~Terashima and M.~Yamazaki, \textit{Semiclassical Analysis of the 3d/3d
  Relation}, {{
  Phys.Rev.} {\bfseries D88} no. 2, (2013) 026011},
{{\ttfamily arXiv:1106.3066}}.

\bibitem[TM3]{Terashima:2012cx}
Y.~Terashima and M.~Yamazaki, \textit{Emergent 3-manifolds from 4d Superconformal
  Indices}, {{
  Phys.Rev.Lett.} {\bfseries 109} (2012) 091602},
{{\ttfamily arXiv:1203.5792}}.

\bibitem[W]{Witten:1982df}
E.~Witten, \textit{Constraints on Supersymmetry Breaking},
{{Nucl.Phys.}
  {\bfseries B202} (1982) 253}.


\end{thebibliography}

\end{document}